\documentclass{PoS}
\usepackage[english]{babel}
\usepackage{subfig}
\usepackage[font=small,labelfont=bf]{caption}
\usepackage[numbers]{natbib}

\title{Muon tagging on the Backend-Electronics of CHEC-S - a compact high-energy camera for the Cherenkov Telescope Array}

\ShortTitle{Muon tagging on the BEE of CHEC-S}

\author{Roberta Pillera$^{a}$, Gianluca Giavitto$^b$, \speaker{Heike Prokoph$^b$} for the CTA Consortium\footnote{for consortium list see PoS(ICRC2019)1177}\\
	 \llap{$^a$} University of Bari and INFN Bari\\
     \llap{$^b$} Deutsches Elektronen Synchrotron (DESY), Zeuthen\\
     
     E-mail: \email{roberta.pillera@ba.infn.it}, \email{gianluca.giavitto@desy.de}, \email{heike.prokoph@desy.de}}
\abstract{The Cherenkov Telescope Array (CTA) will be the leading ground-base observatory for Very High Energy (VHE) $\gamma$-ray astronomy for the next decades. Its southern site will host about 70 Small Sized Telescopes (SSTs) which will determine the CTA sensitivity at $\gamma$-ray energies between ~1 and ~300 TeV. One of the design options for the SST cameras is the silicon photomultiplier-based Compact High-Energy Camera (CHEC-S). The back-end electronics (BEE) of CHEC-S interconnects the camera front-end modules, provides power and clock distribution, aggregation, routing and timestamping of data and most importantly it implements the camera trigger system. A novel technique to tag muons using the capabilities of this system has been developed, studying and comparing different algorithms such as circle fitting, machine learning and simple pixel counting. This contribution describes the design of the CHEC-S BEE, and presents the results of the performance of this muon tagger and the prospects of using it for other Cherenkov Telescopes types of CTA.}

\FullConference{36th International Cosmic Ray Conference -ICRC2019-\\
		July 24th - August 1st, 2019\\
		Madison, WI, U.S.A.}

\begin{document}

\section{Introduction}
The Cherenkov Telescope Array, CTA, will be the major global observatory for very high energy (VHE) $\gamma$-ray astronomy over the next decade and beyond \cite{cta}. It will cover a photon energy range from 20 GeV to 300 TeV. The observatory will operate arrays on sites in both hemispheres to provide full sky coverage and will hence maximize the potential for investigating very energetic non-thermal cosmic accelerators such as supernova remnants, gamma-ray bursts or gravitational wave transients. With 99 telescopes on the southern site and 19 telescopes on the northern site, flexible operation will be possible, with sub-arrays available for specific tasks. There will be telescopes with three different sizes designed to cover the energy range of CTA: the Small Sized Telescopes (SSTs) will have a primary mirror diameter of nearly 4 meters, the Medium Sized Telescope (MSTs) with a primary mirror diameter of 12 meters and the Large Sized Telescopes (LSTs) that will be the largest telescopes, with a diameter of the primary mirror diameter of 23 meters.

One of the challenges in reconstructing the energy of a VHE gamma ray as recorded by an imaging Atmosperhic Cherenkov Telescope (IACT) array such as the future CTA, is constantly monitoring the light throughput of the optical elements of its telescopes. The most straightforward method to achieve this is to record Cherenkov light from muons. Muons produced in Extensive Air Showers (EAS) generate ring-like images in IACTs when travelling nearly parallel to the optical axis. From geometrical parameters of these images, the absolute amount of light emitted may be calculated analytically \cite{vacanti}. Comparing the amount of light recorded in these images to expectation is a well-established technique for telescope optical efficiency calibration \cite{vacanti}. However IACT arrays typically only record events where at least two neighbouring telescopes triggered in coincidence, a technique which suppresses spurious events due to noise, but also local muons. Therefore, one of the requirements for the proposed cameras of CTA is to be able to self-trigger and record muons with high efficiency. This work aims to allow one of the proposed cameras for the small-size telescopes (SSTs) of CTA, CHEC-S, to trigger on muons. 

\section{Back-end electronics of CHEC-S}
\begin{figure}
	\centering
	\begin{minipage}[b]{6cm}
	\centering
	\includegraphics[width=6cm]{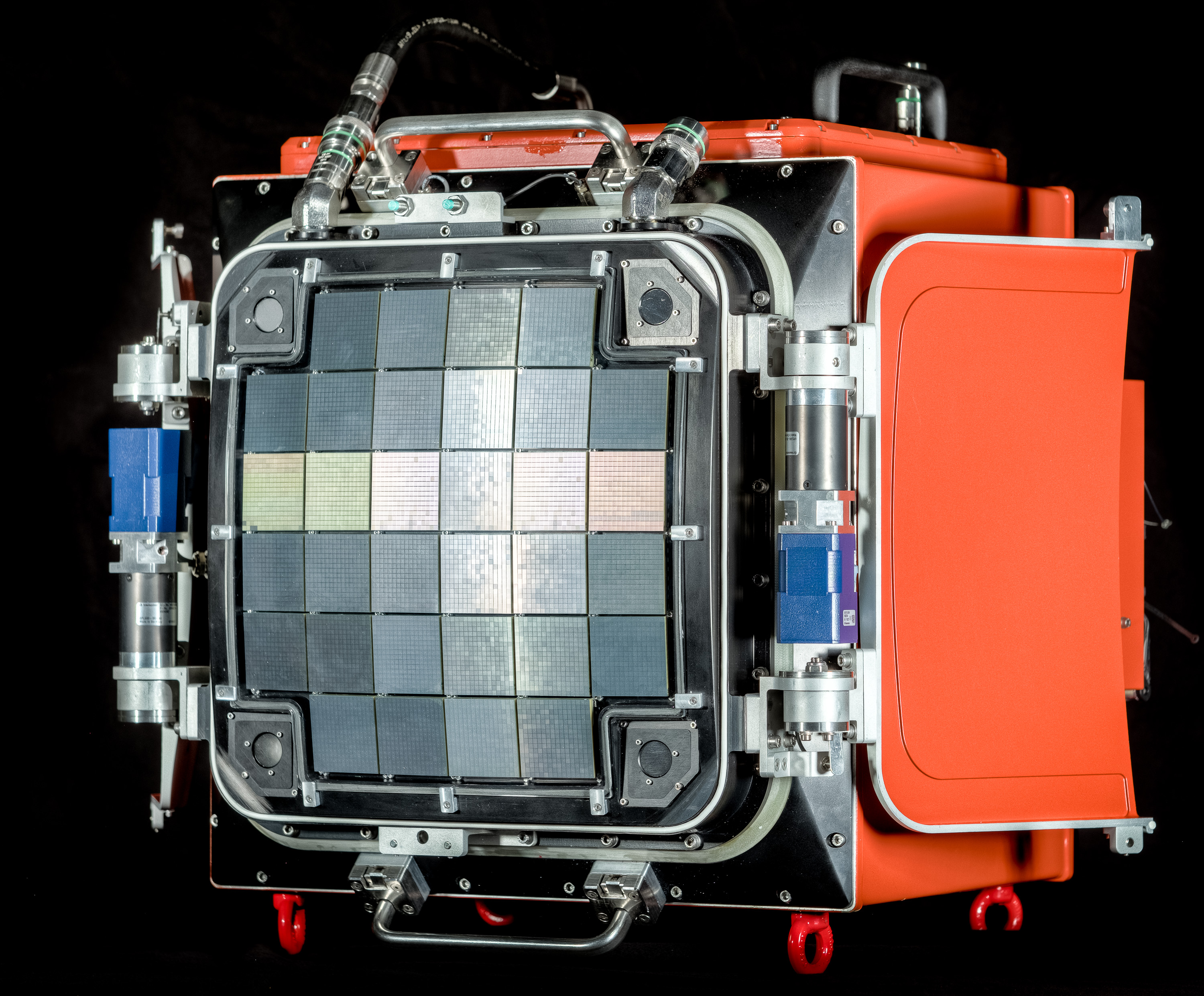}
	\caption*{a}
	\end{minipage}
	%\hspace{10mm}
	\begin{minipage}[b]{60mm}
	\centering
	\includegraphics[width=60mm]{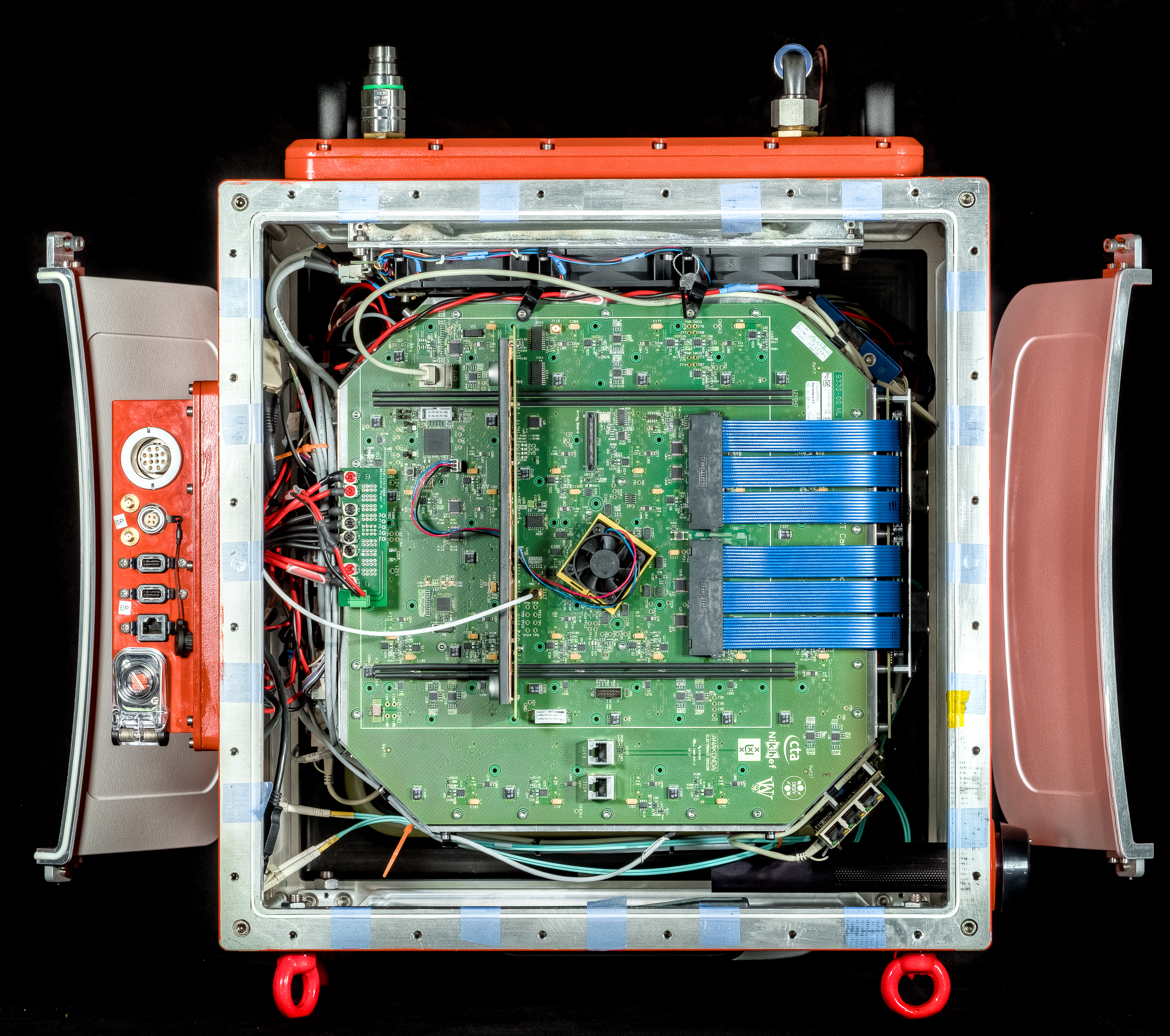}
	\caption*{b}
	\end{minipage}
\caption{a): Front of CHEC-S, showing the SiPM focal plane array b) Back of CHEC-S with back plate removed, showing the backplane.}
\label{chec_images}
\end{figure}

CHEC-S \cite{chec} is a Silicon photomultiplier-based camera designed for Schwarzschild-Couder double-mirror telescopes. It is designed to be cost-effective, using commercial off-the-shelf components where possible. It has 2048 6x6mm pixels divided in 32 tiles of 8x8 pixels each (see Fig. \ref{chec_images}a). Each tile is connected to a front-end buffer, then to a module providing readout and triggering capabilities. The trigger is formed in a dedicated ASIC (T5TEA, derived from TARGET 5 \cite{target}) by summing up the signal of groups of 4 adjacent pixels, called superpixels, and comparing it to a trigger threshold. Thus, the camera front-end delivers 512 independent superpixel trigger signals to the CHEC-S backplane.

The CHEC-S backplane is a large (50x50 cm) 16-layer board serving several purposes, the most important of which is generating a camera trigger using these 512 superpixel signals (see Fig \ref{chec_images}b). These are isochronally routed from each module connector to a large Xilinx Virtex 6 FPGA. There, they are sampled at 1 GHz and a neighbouring logic is applied: the camera backplane issues a trigger whenever two neighbouring pixels in the horizontal, vertical or diagonal direction trigger at the same time with a coincidence window of ~2.7 ns. Simultaneously, the FPGA records the 512-bit superpixel trigger patterns and sends them to the central camera server as a UDP packet, together with timestamping information. These patterns are in practice low-resolution, B/W shower images. The goal of this study was to develop a way to tag muons efficiently and rapidly using the information contained in these images, with enough performance to meet the CTA requirements.

\section{Muon tagging}
\begin{figure}
	\centering
	\begin{minipage}[b]{10cm}
	\centering
	\includegraphics[width=10cm]{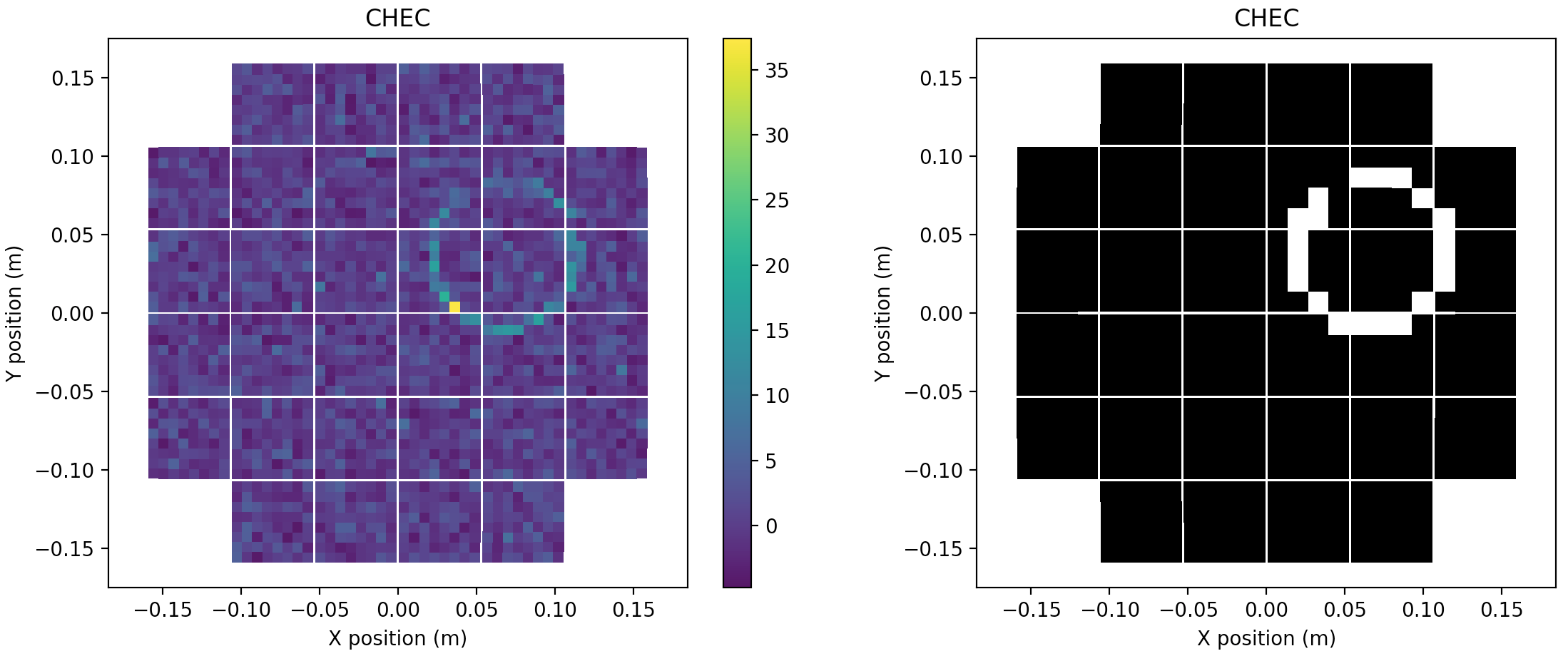}
	\caption*{a}
	\end{minipage}
	\hspace{10mm}
	\begin{minipage}[b]{10cm}
	\centering
	\includegraphics[width=10cm]{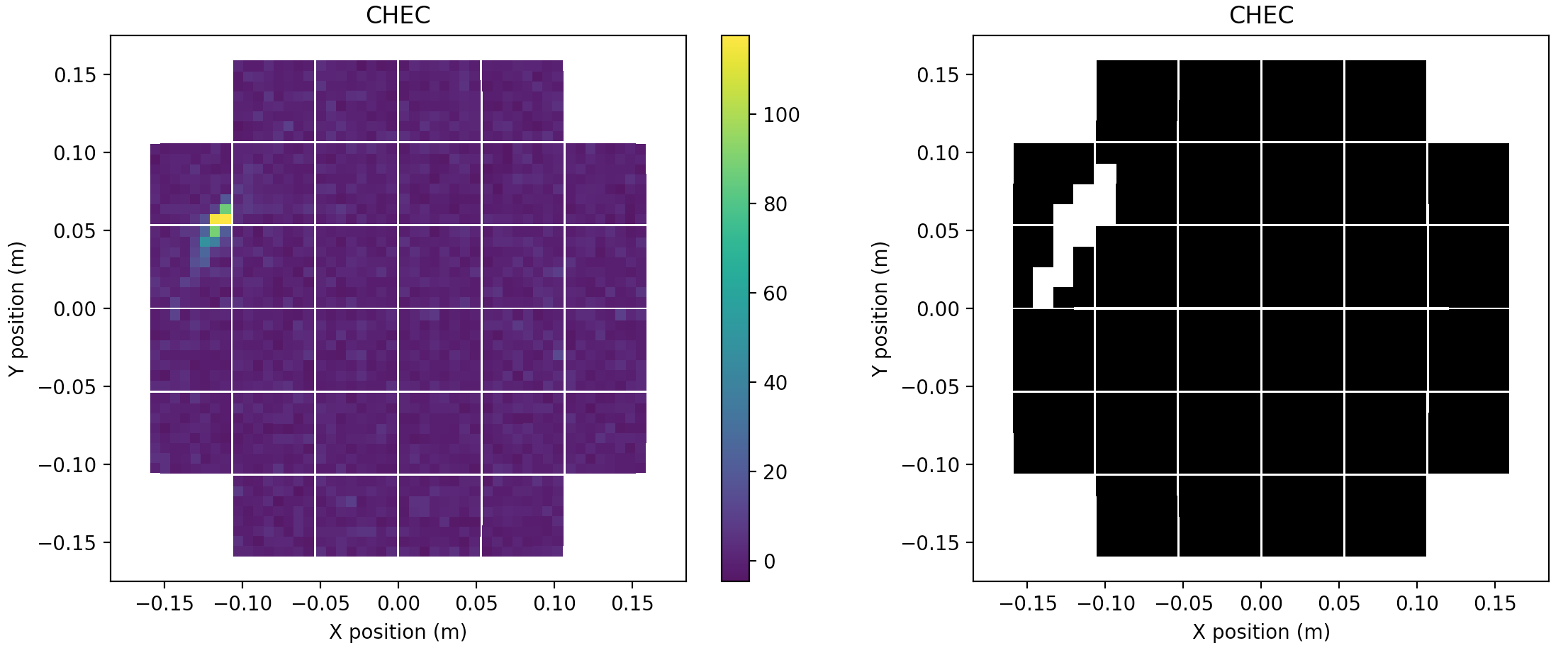}
	\caption*{b}
	\end{minipage}
\caption{Simulation of a 22.69 GeV muon a) and a 5.41 TeV proton shower b) with Cherenkov image (left) and triggerpattern image (right).}
\label{comparison_pic}
\end{figure}

The main background for muon identification are the proton shower images. We investigated different algorithms fast enough to operate on trigger info at the camera server level. The requirement placed by CTA is to flag fully contained muon rings with an overall efficiency above 90 \%. Examples of simulated muon and proton shower images are shown in Fig. \ref{comparison_pic}. On the left there is the full recorded Cherenkov image and on the right there is the superpixel trigger data image.
A preliminary study on different algorithms for muon/proton discrimination was performed testing speed and efficiency. All of them rely on the charactersitic circular shape of the muon image. We investigated a circle fit procedure with the Taubin method \cite{astri} to evaluate the geometrical parameters of the ring as center and radius. The centroid of the image is given as initial guess to the Taubin minimization function. Another approach that we studied was machine learning image recognition using a neural network (Multilayer Perceptron). The last method we tested was Majority, that consists in setting a discrimination threshold on the number of fired superpixels. The event processing rates were evaluated for all methods and are summarized in Tab. \ref{tab1}.
\begin{table}[h]
\centering
\begin{tabular}{c||c} 
	method & rate\\
	\hline
	Taubin & 188 Hz\\
	MLP & 63 kHz\\
	Majority & 130 kHz
\end{tabular}
\caption{Average event processing speed of each method at 95\% efficiency.}
\label{tab1}
\end{table}
In order to match the CHEC-S readout rate we require the method to be faster than 1 kHz. The Taubin fitting method does not fulfill this requirement, while the other two show good and similar performance. The fitting algorithm could be improved to speed up computation time, but probably without reaching the performances of the other two algorithms. The simplicity of the Majority method could allow the development of fast trigger algorithm to be implemented directly in the backplane of the camera, hence detailed studies were performed on it. The distributions of the muon signal and the proton shower background of the Majority classifier are shown in Fig. \ref{distr_plots}a).
\subsection{Stability in terms of Night Sky Background}
\begin{figure}
	\centering
	\begin{minipage}[b]{6cm}
	\centering
	\includegraphics[width=6cm]{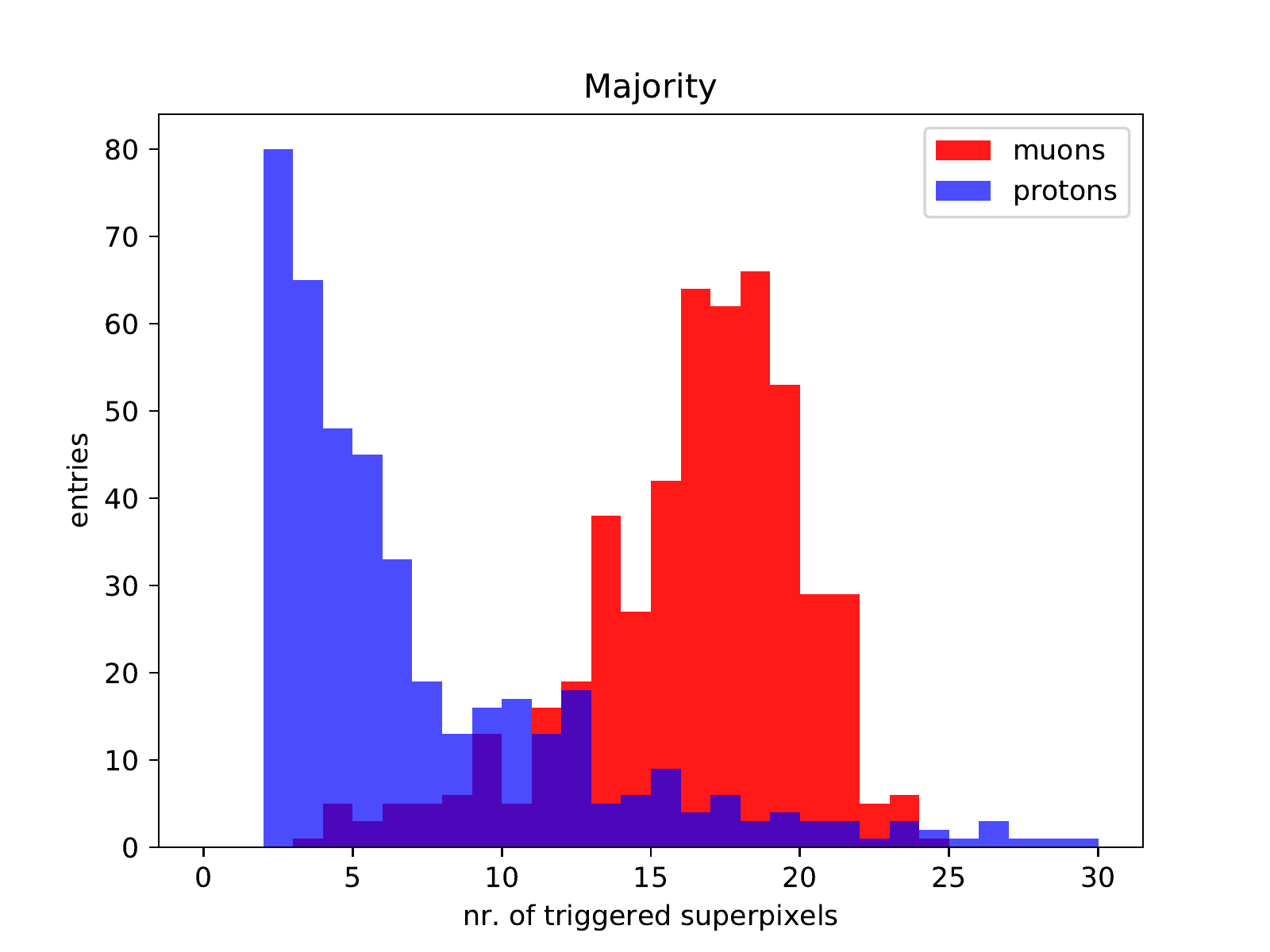}
	\caption*{a}
	\end{minipage}
	%\hspace{10mm}
	\begin{minipage}[b]{60mm}
	\centering
	\includegraphics[width=60mm]{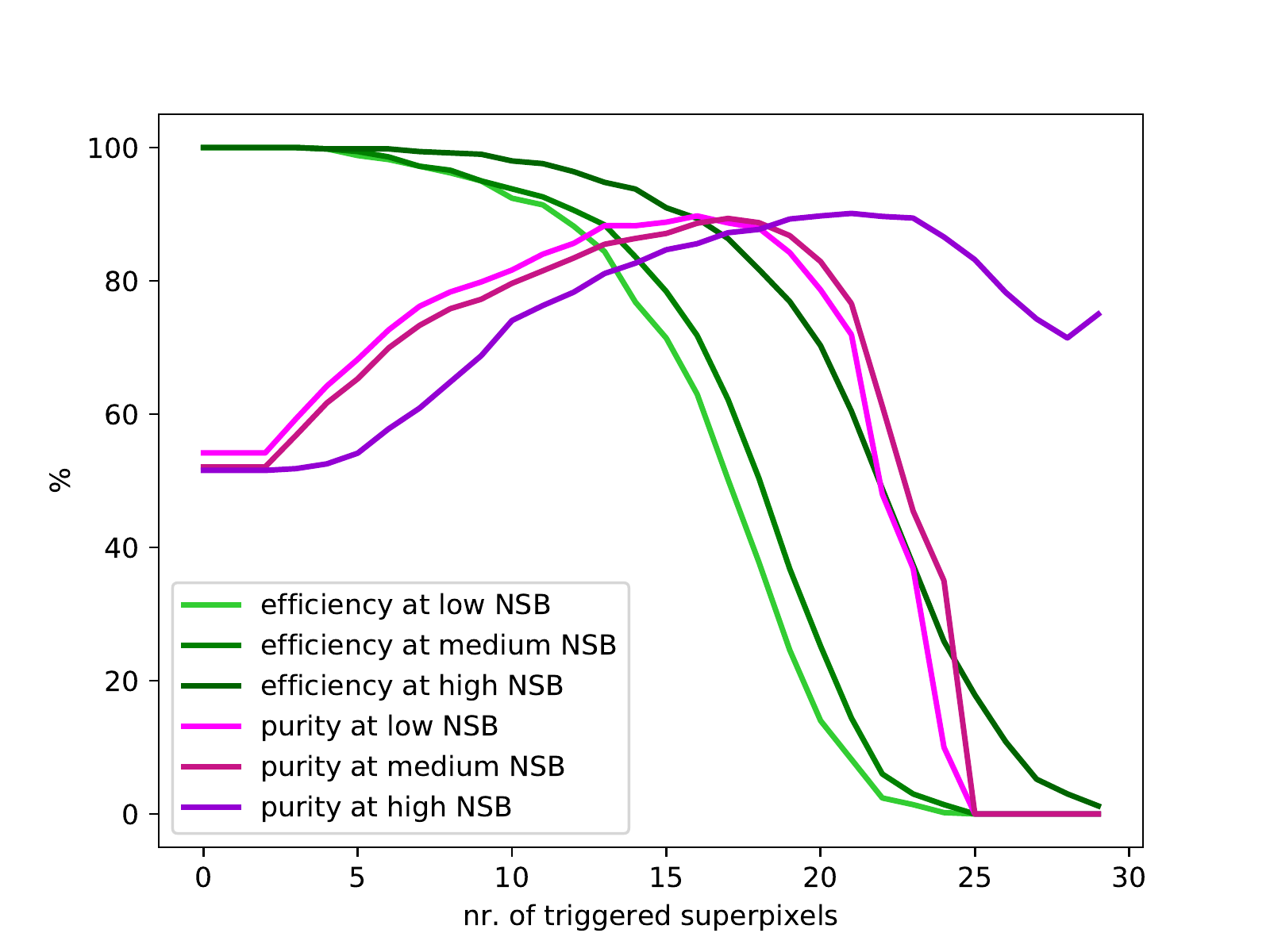}
	\caption*{b}
	\end{minipage}
\caption{a): Superpixel distributions of muon signal and proton shower background with low NSB. b) Selection efficiency and purity plots for different NSB levels. Muons were simulated in an energy range from 8 GeV up to 1 TeV and protons from 1 to 300 TeV, both with a spectral index of $\gamma = -2.0$.}
\label{distr_plots}
\end{figure}
An important aspect that has to be considered is the stability with changing Night Sky Background (NSB). Different observation regions correspond to different NSB. Observations in the extragalactic field will have a low NSB, and this is what is considered in the standard simulations used above. Observations on the galactic plane and the galactic center correspond respectively to medium and high NSB. This can affect the selection efficiency. However studies on simulations with different NSB levels (Fig. \ref{distr_plots}b) show that fixing a threshold that provides 90 \% efficiency at low NSB, the overall efficiency of the Majority remains above 90 \% even at increasing NSB.
\section{Conclusions}
In this contribution we have presented the current status of the muon tagging algorithms study for a muon trigger and showed the first preliminary results obtained with simulations. We investigated different methods to tag muons from trigger pattern images and we found that the Majority method seems to have the best performance in terms of speed, background rejection and stability to increasing NSB. A deeper study of the efficiency of the Majority tagger as well as a study on its performance in non-ideal conditions is ongoing and the application on the backend electronics of CHEC-S is foreseen.

\end{document}